%% file: main.tex
\documentclass[conference]{IEEEtran}
\usepackage{cite}
\usepackage{amsmath,amssymb,amsfonts}
\usepackage{algorithmic}
\usepackage{graphicx}
\usepackage{textcomp}
\usepackage{xcolor}
\usepackage[hyphens]{url}

\usepackage{url}
\usepackage{makecell}
\usepackage{comment}
\usepackage{colortbl}
\usepackage{balance}
\usepackage{xspace}
\usepackage{caption}
\usepackage{subcaption}

\def\BibTeX{{\rm B\kern-.05em{\sc i\kern-.025em b}\kern-.08em
    T\kern-.1667em\lower.7ex\hbox{E}\kern-.125emX}}

\title{WaSP: Warp Scheduling to Mimic Prefetching in Graphics Workloads }

\author{\IEEEauthorblockN{Diya Joseph\IEEEauthorrefmark{1},
Juan L. Aragón\IEEEauthorrefmark{2},
Joan-Manuel Parcerisa\IEEEauthorrefmark{1} and
Antonio González\IEEEauthorrefmark{1}}
\IEEEauthorblockA{\IEEEauthorrefmark{1} 
Universitat Politècnica de Catalunya,
Barcelona, Spain\\}
\IEEEauthorblockA{\IEEEauthorrefmark{2} 
Universidad de Murcia,
Murcia, Spain\\}}

\begin{document}
\maketitle
\thispagestyle{plain}
\pagestyle{plain}

\input{0_Abstract/Abstract}
\input{1_Introduction/Introduction}
\input{2_Background/Background}

\input{3_Technique/Technique}

\input{4_Methodology/Methodology}
\input{5_Evaluation/Evaluation}
\input{6_Related_Work/Related_work}

\input{7_Conclusions/Conclusions}

\bibliographystyle{IEEEtranS}
\bibliography{refs}

\end{document}

%% file: 0_Abstract/Abstract.tex
\begin{abstract}
Contemporary GPUs are designed to handle long-latency operations effectively; however, challenges such as core occupancy (number of warps in a core) and pipeline width can impede their latency management. This is particularly evident in Tile-Based Rendering (TBR) GPUs, where core occupancy remains low for extended durations. To address this challenge, we introduce WaSP, a lightweight warp scheduler tailored for GPUs in graphics applications. WaSP strategically mimics prefetching by initiating a select subset of warps, termed \textit{priority warps}, early in execution to reduce memory latency for subsequent warps. This optimization taps into the inherent but underutilized memory parallelism within the GPU core. This underutilization is a consequence of a baseline scheduler that evenly spaces misses throughout execution to exploit the inherent spatial locality in graphics workloads. WaSP improves on this by reducing average memory latency while maintaining locality for the majority of warps. While maximizing memory parallelism utilization, WaSP prevents saturating the caches with misses to avoid filling up the MSHRs (Miss Status Holding Registers). This approach reduces cache stalls that halt further accesses to the cache. Overall, WaSP yields a significant 3.9\% performance speedup. Importantly, WaSP accomplishes these enhancements with a negligible overhead, positioning it as a promising solution for enhancing the efficiency of GPUs in managing latency challenges.
\end{abstract}

%% file: 1_Introduction/Introduction.tex
\section{Introduction}
\label{sec:introduction}
Graphics Processing Units (GPUs) are specifically designed to handle tasks involving long-latency operations. Despite their efficiency, challenges arise due to factors like warp occupancy within the core and pipeline width, impacting the GPU's ability to conceal latency. The widening gap in frequency between GPUs and memory results in increased latencies and thus aggravates these challenges, particularly in Tile-Based Rendering (TBR) mobile GPUs, where low core occupancy persists for extended periods.

To address this challenge, a possible approach involves scaling up the number of warps concurrently processed within a GPU core. However, this requires amplifying all resources needed for each warp, notably expanding the register file size, which directly scales with the warp count. This solution incurs substantial costs. Alternatively, our proposal aims to tackle increased latencies by directly mitigating latency itself rather than opting for an expensive resource-intensive solution.

In TBR GPUs, frames are processed in square subsets called tiles, with each warp assigned the rendering of a fixed number of pixels. Warps in graphics workloads exhibit spatial and temporal locality when accessing textures due to the proximity of pixels. Warps with pixels in close proximity access texels that are close to one another, or even the exact same texels sometimes. Typically, warps are scheduled in the shader cores (GPU cores) in a scanline order of pixels. This leads to proximate warps becoming consecutive warps with high cache locality. 

\begin{figure}[t]
              \centering
              \includegraphics[width=\linewidth]{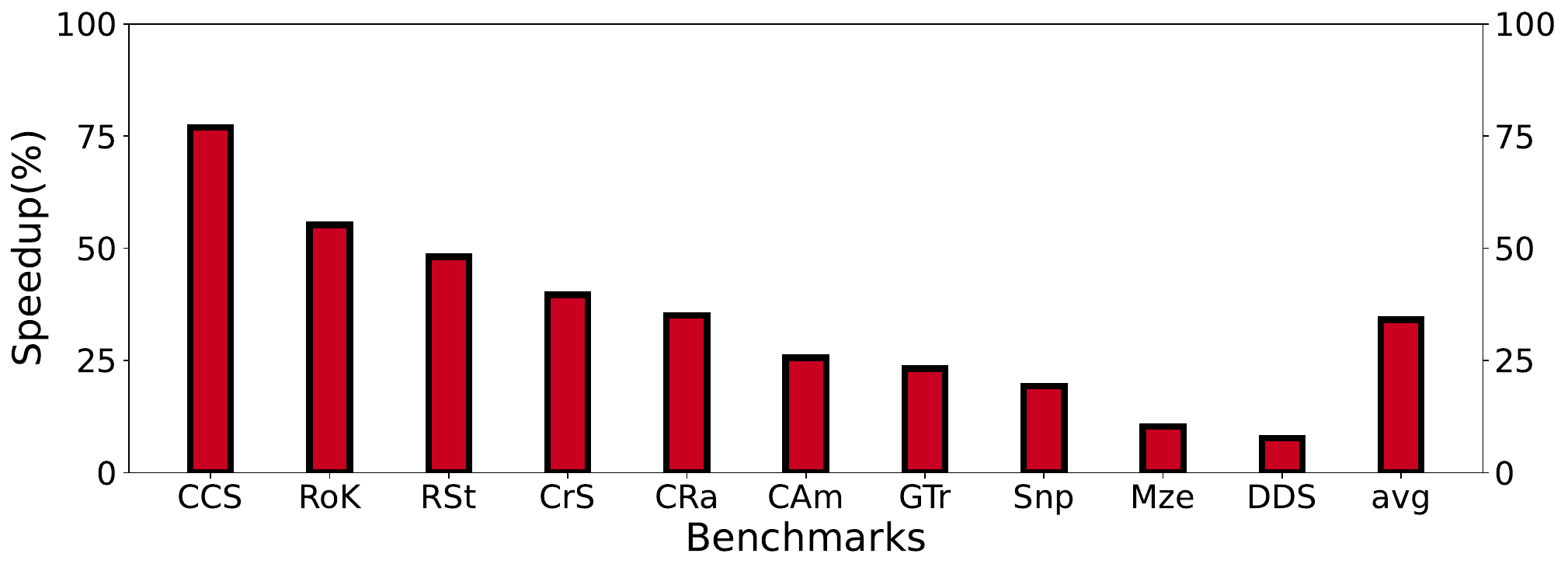}
              \caption{Speedup w.r.t. an ideal main memory with zero latency.}
              \label{fig:motivation}
\end{figure}

Although there is natural cache locality resulting from this arrangement, it does not fully exploit the memory parallelism offered by the memory hierarchy. We will show that some long-latency misses issued to main memory are not well overlapped, leading to inefficient latency hiding mechanisms. Figure \ref{fig:motivation} shows the potential speedup of real-world applications in our benchmark suite when using an ideal main memory with zero latency. Thus, we see that graphics applications in TBR GPUs show promise in speedup with a reduced memory latency.

In response to this challenge, we introduce WaSP, a novel scheduler designed to prioritize a specific subset of warps, referred to as \textit{priority warps}, for execution ahead of others. By effectively emulating prefetching for the remaining warps, WaSP significantly reduces the average memory latency experienced by each warp, leading to a latency reduction of 9\%. This results in a speedup of 3.9\%, while incurring minimal hardware overhead.

WaSP optimizes this process by isolating a small subset of warps that accurately covers the majority of texture memory blocks accessed in the tile, thus maximizing memory-level parallelism (MLP). Known as ``priority warps'', these warps are strategically allocated to GPU cores until the memory unit is about to be saturated with outstanding misses. Subsequently, regular warps are scheduled until the LDST unit gains more availability, allowing for the prioritization of the ``priority warps'' once again. This method ensures a more efficient utilization of the memory resources, reducing the average miss latency for memory accesses across the regular warps. By employing this approach, WaSP better exploits memory-level parallelism while carefully avoiding blocking the caches in the memory hierarchy, a phenomenon detailed in Section \ref{subsec:cachestalls}.

\begin{figure*}
            \centering
            \includegraphics[width=\linewidth]{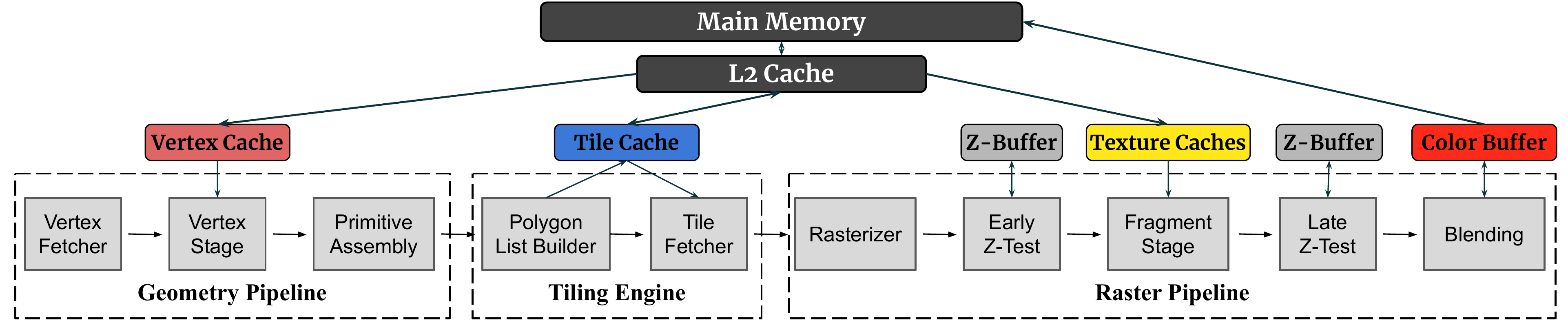}
            \caption{The Graphics Pipeline of a TBR GPU.}
            \label{fig:Graphics_Pipeline}
    \end{figure*}

To summarize, this paper makes the following key contributions:
    \begin{itemize}
        \item \textbf{Priority Warp Selection}: Proposes and evaluates various subsets of warps as priority warps to maximize the coverage of the full texture footprint of a tile.
        
        \item \textbf{Priority Warp Scheduling}: Introduces and evaluates different heuristics for transitioning between priority and regular warps, preventing the blocking of the memory unit thanks to using priority warps thus leading to an overall performance gain.
        
        \item \textbf{Performance Improvements}: Demonstrates a 3.9\% increase in IPC on average with minimal hardware overhead.
        
    \end{itemize}

The rest of this paper is organized as follows: Section 2 provides background information on GPUs, with a special emphasis on texture locality and warp scheduling. Section 3 presents a detailed description of WaSP. The tools and workloads used to evaluate our technique are described in Section 4. Section 5 presents our experimental results and analysis. Section 6 reviews related work, and Section 7 summarizes the main conclusions of the paper.

%% file: 2_Background/Background.tex
\section{Background}
    \label{sec:background}    

    Mobile GPUs typically employ a Tile-Based Rendering (TBR) architecture. The idea for TBR architectures was initially proposed to enable parallel rendering \cite{fuchs1989pixel}, \cite{molnar1994sorting}. \textit{Tiles} are disjoint segments of the frame that can be rendered in parallel. Nowadays, TBR is a common architecture adapted for low-power graphics systems where, instead of tiles being rendered in parallel, they are rendered sequentially over small tile-sized on-chip buffers. This approach capitalizes on locality, minimizing power-intensive DRAM accesses and conserving memory bandwidth. According to a work by Antochi \textit{et al.} \cite{antochi2004memory}, a TBR architecture reduces the total amount of external data traffic by a factor of 1.96 compared to a GPU architecture that is not tile-based (a.k.a. Immediate Mode Rendering).
    
    \subsection{Graphics Pipeline}
    \label{subsec:Graphics_Pipeline}    
        Figure~\ref{fig:Graphics_Pipeline} illustrates the key stages of the Graphics Pipeline and provides an overview of the memory hierarchy organization.
        In Raster Graphics Systems, the Geometry Pipeline transforms the geometric description of a scene and creates all the \textit{primitives} that fall inside the frustum view in accordance with the camera's viewpoint. On the other hand, the Raster Pipeline discretizes each primitive into \textit{fragments} (at pixel granularity) that are then shaded and blended to produce the final screen image.
    
        In a TBR architecture, the Raster Pipeline is designed to render \textit{tiles} rather than the full frame. These tiles are usually square groups of adjacent pixels. This tiling improves locality and allows keeping on chip most bandwidth-intensive memory accesses. In order for this to happen, all the geometry needs to be sorted into subsets that will individually be able to fully render the image for each of these tiles\cite{molnar1994sorting}. The process of tiling is carried out by a new pipeline stage called \textit{Tiling Engine}.
    
        Thus, the Graphics Pipeline for TBR architectures consists of three parts, namely the Geometry Pipeline, the Tiling Engine and the Raster Pipeline, as shown in Figure \ref{fig:Graphics_Pipeline}.
        
        Input data for the Graphics Pipeline consists of vertices and textures. These vertices join to form different polygons (usually triangles) called \textit{primitives}  and the textures are used to enhance details on surfaces while rendering the scene. 
        
        A \textit{Draw Command} triggers the Geometry Pipeline and the Vertex Stage starts fetching vertices from memory using an L1 Vertex Cache. It then transforms them according to a vertex program provided by the user. The Primitive Assembler takes the vertices in program order and joins them to produce primitives. These primitives are fed as input to the Tiling Engine.
        
        The goal of the Polygon List Builder is to produce a list, for each tile of the screen, containing all the primitives that overlap it. This data is arranged in a structure known as the \textit{Parameter Buffer}. 
        
        After all the geometry is processed and binned, the Tile Fetcher fetches the primitives corresponding to each tile in the frame, one tile at a time. Tiles are processed in an order specified by the Tiling Engine, and their primitives are put into a FIFO queue for the Raster Pipeline to consume.
        
        The Raster Pipeline renders each tile sequentially. For this purpose, the Rasterizer takes each primitive from the FIFO queue and identifies which pixels of the current tile are overlapped by the primitive. It then uses interpolation to calculate attributes for each pixel, a set of data called \textit{fragment}. The fragments of every four adjacent pixels are grouped to form a \textit{quad}, and these quads are sent to the Early Z-Test stage. This stage uses a tile-sized buffer called the \textit{Z-Buffer} to store the minimum depth of previously processed fragments on each tile’s pixel coordinate in order to eliminate those that lie behind another previously processed opaque fragment. The non-discarded quads are then sent to a GPU core, which computes an initial color for each pixel of a quad, taking into account the lighting and textures provided by the \textit{shader program}. The output colors are then sent to the Blending Unit. This unit computes the final color of pixels depending on the transparency of each quad, and stores them in the Color Buffer. Some rendering techniques require that the GPU core changes the depth of fragments, in which case the Early Z-Test is disabled and the Late Z-Test is employed. Note that both the Color Buffer and the Z-Buffer have the size of just one tile, and thus can be stored on-chip. Finally, the Color Buffer is flushed to the Frame Buffer in main memory, after a tile has been completely processed. Quads are propagated between stages through FIFO queues.

        \begin{figure}[t]
              \centering
              \includegraphics[width=\linewidth]{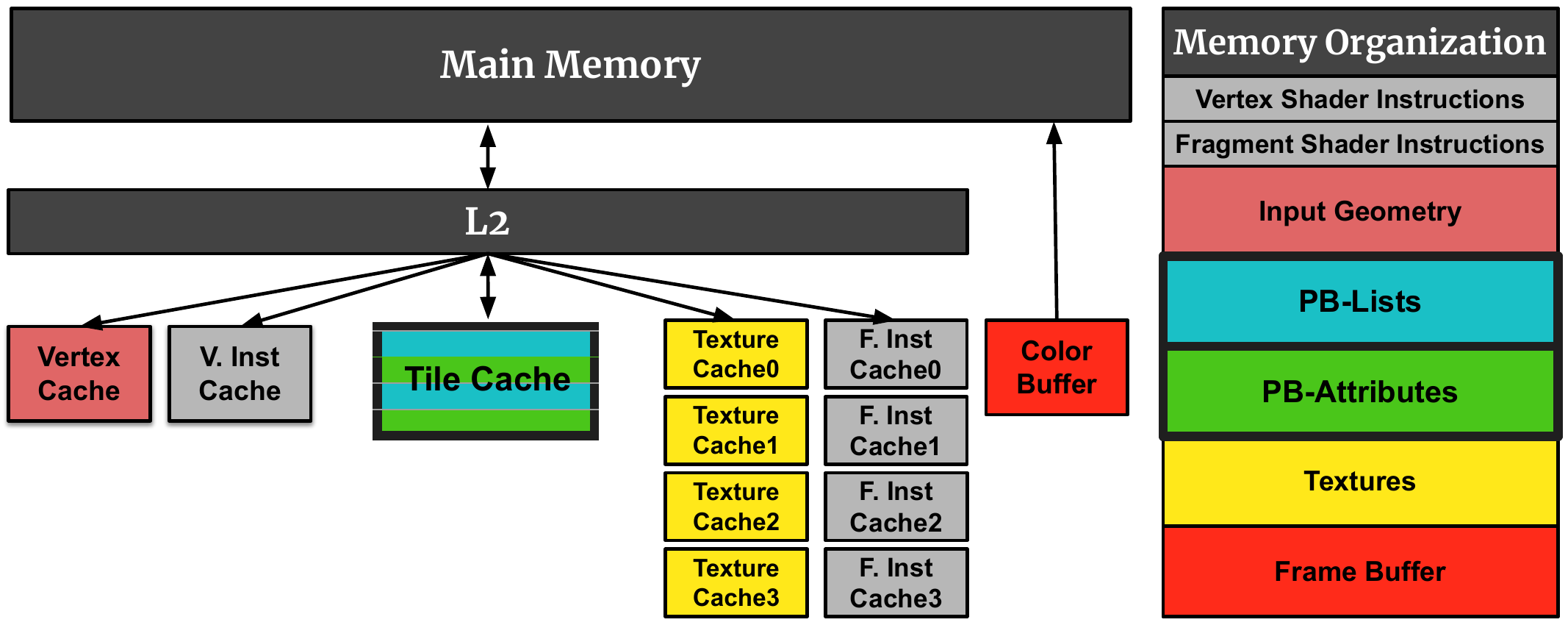}
              \caption{Baseline memory hierarchy and memory organization.}
              \label{fig:Baseline Mem Hierarchy and Mem Organisation}
            \end{figure}

    \subsection{Memory Organization}
            Figure \ref{fig:Baseline Mem Hierarchy and Mem Organisation} illustrates the primary memory data structures of a graphics application (on the right) alongside the memory hierarchy utilized for their storage and access (on the left). Notably, there are multiple L1 caches dedicated to instructions and data, supported by a shared L2 cache, ultimately backed by main memory. Also note that in case of larger pipeline widths (like 4) we add an 0.5 KB L0 with two ports to each GPU core. This is done to meet the increased bandwidth requirement of that wider pipeline. We find that a small sized L0 works because of the inherent locality of graphics workloads, as explained below.
       
    \subsection{Texture Locality}
        \label{subsec:texlocality}
            To comprehend the impact of warp order on the locality and memory parallelism of memory accesses, it is crucial to grasp the access pattern of the data within the GPU core. This data primarily comprises textures.
            
            Textures function as images or patterns applied to 3D models or 2D surfaces, enriching their visual appearance by offering intricate surface details such as color, reflectivity, and roughness. These visual cues contribute to making graphical objects appear more lifelike and realistic.
                        
            When incorporated into a graphics application, textures must be stored in the computer's memory for quick access during rendering. Texture locality patterns play a vital role in optimizing rendering performance. These patterns determine how efficiently texture data can be accessed and utilized by the graphics hardware within a frame.
            
            There are two key patterns of texture locality recognized:
            
            \begin{itemize}
                \item \textbf{Spatial Locality}: This pattern refers to nearby pixels in an image accessing similar or neighboring texels. A ``texel'' represents a texture element, the fundamental unit of a texture map. The number of texels required for texture mapping a single pixel varies based on the filtering technique employed by the programmer. For instance, bilinear filtering requires 4 adjacent texels, while trilinear filtering requires 8 adjacent texels. To exploit spatial locality effectively, texels are often stored in a way that maximizes neighboring texels to be stored in the same memory block.
                
                \item \textbf{Temporal Locality}: The set of texels accessed by a pixel often overlaps with the set of texels of the neighboring pixels. 
            \end{itemize}
            
            In summary, recognizing and capitalizing on the spatial and temporal locality of adjacent pixels underscores the importance of strategically scheduling neighboring warps into GPU cores. This scheduling plays a vital role in maximizing the efficient use of both temporal and spatial locality for optimized performance.
        
    \subsection{Warp Scheduling}
    \label{subsec:warpscheduling}
            Four fragments are grouped into a quad, and a quad is launched as a warp in the GPU core. The shader program, referred to in Section \ref{subsec:Graphics_Pipeline}, corresponds to a kernel. In the baseline scheduling approach, warps are sequenced in the same order as they are rasterized in the Rasterizer, which follows a scanline pattern. This method ensures that adjacent quads, spatially close to each other, are executed in proximity. Consequently, the spatial locality within textures is effectively exploited.
            
            However, this approach poses a challenge. Prior to the commencement of execution in a GPU Core, the working set of a tile for textures is not yet present in the L2 or L1 caches. Typically, the working set of a tile fits comfortably within these caches, allowing all but the first accesses to result in cache hits or secondary misses. Given the scanline-based scheduling, the primary (first) misses to a memory block are evenly distributed across the tile. This is because scanline ensures that adjacent warps have high locality and thus unique memory block accesses are far apart from each other. Consequently, these evenly spaced cache misses do not overlap well, leading to inefficient utilization of the memory parallelism offered by the memory hierarchy. This effect is explained with an example in Section \ref{subsec:example}.
            
            All warps in a tile need to finalize before another tile starts its execution in the GPU core. Thus, we have many occurrences of filling the core and draining the core with warps. These periods have low core occupancy. Thus hiding long-latency operations, such as these cache misses, is a challenging task in TBR GPUs. Therefore, the primary objective of this work is to reduce latency by exploiting this underutilized memory parallelism effectively. By enhancing core performance through the optimized utilization of memory parallelism, we aim to address these issues and improve the overall efficiency of the GPU.

    \subsection{Cache Stalls} 
    \label{subsec:cachestalls}
        The provisioned number of Miss Status Holding Registers (MSHRs) is determined based on the capacity needed to retain outstanding misses within a cache. The count of outstanding misses at any given time relies on the miss latency and the bandwidth of the cache's connection to upper levels. The bandwidth and the latency varies based on how far the miss travels in the memory hierarchy, making it application-specific.
        
        Greater miss occurrences in the upper levels require a higher count of MSHRs in the lower levels like the L0 and L1. Usually, the MSHRs in the L0 and L1 are provisioned assuming a high hit rate in the next cache level. So, in burst modes where these long-latency misses are closely clustered, all MSHRs can quickly fill up for caches like the L0 and L1. Consequently, the cache reaches a point where it cannot accommodate further accesses since there is no space available to store additional misses resulting from those accesses. Note that this also stalls potential hits in the cache. This condition, known as a cache stall, can have a major impact on performance.

%% file: 3_Technique/Technique.tex
\section{WaSP}
\label{sec:technique}
      The main contribution of this work is to propose a lightweight warp scheduler for GPUs for graphics applications. The key objective is to populate the L1 and L2 caches with the tile's working set early in the tile's execution using a small subset of warps (called \textit{priority warps}), aiming to minimize memory latency for subsequent warps (called \textit{regular warps}) within the tile. In other words, priority warps are used to prefetch data for regular warps. This is done by harnessing the existing underutilized memory parallelism at the beginning of the tile without blocking the memory unit due to cache stalling. This ensures that performance is not only maintained but also enhanced in the process.
      
      The subset of priority warps are chosen to have a high coverage of unique memory block accesses. Ideally, the memory footprint of the priority subset should be the same as that of the whole set of warps in the tile. This is hard to achieve so we try to be as close as possible.

      The challenges involved in proposing such a scheduler revolve around two key aspects:

      \begin{itemize}
            \item \textbf{Priority Warp Selection}: The selection process for the subset of priority warps focuses on achieving a high coverage of unique memory block accesses. The goal is to match the memory footprint of the priority subset with that of the entire set of warps in the tile while keeping the subset size relatively small. This task is non-trivial but essential to ensure that optimizations applied to these representative warps significantly impact overall performance.

            \item \textbf{Priority Warp Scheduling}: It is crucial to prevent scenarios where priority warps lead to blocking the memory unit, and consequently also the pipeline, due to filling up the MSHRs in L0 or L1 caches. These events are particularly likely, since the priority subset contains a higher density of unique memory block accesses, which increases the risk of memory unit blocking. These stalls could have more severe consequences than inefficient latency overlapping, as they may also result in delays for hits.
      \end{itemize}
      
      Before diving into a detailed explanation of our proposal, let us first consider an example to grasp the problem and its solution at a high level.
      
      \input{3_Technique/2_Example}
      \input{3_Technique/3_PW_selection}

      \input{3_Technique/4_PW_timing}

      \input{3_Technique/5_HW_overhead}

%% file: 3_Technique/2_Example.tex
\subsection{The Example}
    \label{subsec:example}
    \begin{figure}[t]
              \centering
              \includegraphics[width=\linewidth]{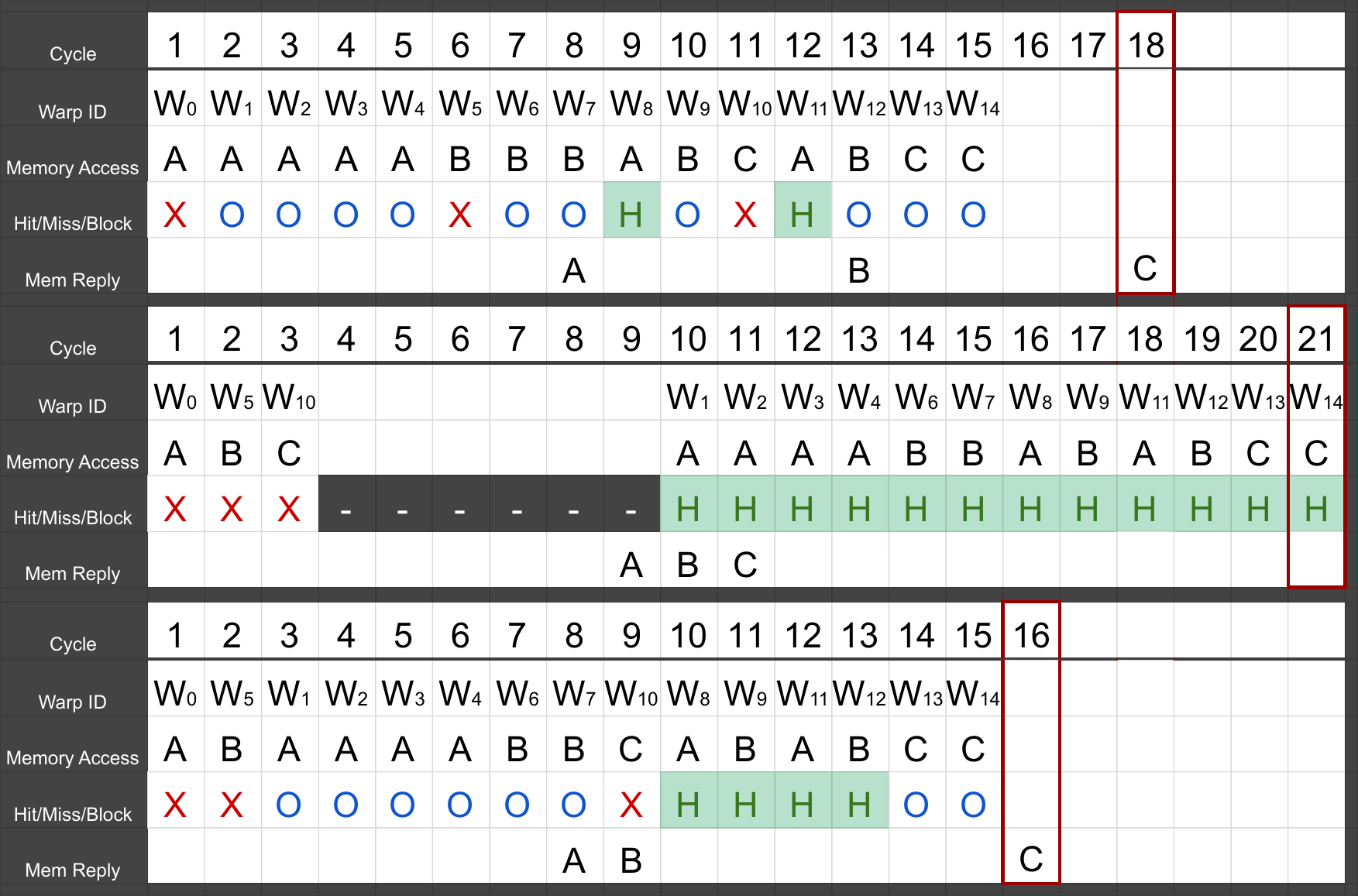}
              \caption{The Example.}
              \label{fig:Example}
    \end{figure}
    Consider a tile with 15 warps, each generating a single memory access, that collectively access to three unique memory blocks—A, B, and C. We assume a main memory with a fixed latency of 7 cycles and a single-level cache with an MSHR size of 3. While the order of warps can influence hits and misses, our analysis assumes that once fetched from main memory, memory blocks remain available in the lower memory levels until the tile's completion. This is a reasonable assumption given that working sets of tiles typically fit comfortably in L1 and L2 caches. Thus, the order of warps does not change the number of primary misses. Out of the fifteen accesses, three result in primary misses to main memory, while the rest result in hits or secondary misses.
    
    Figure \ref{fig:Example} includes three chronographs illustrating four aspects. First, the memory block accessed in a cycle, and second, its corresponding Warp ID. Third, the cache status—whether it yielded a hit, miss, or secondary miss. In our representation, X denotes a miss, O denotes a secondary miss, and H denotes a hit. The dark blank cycles indicate that the memory is blocked because the MSHR is fully occupied by three misses, causing a blockage in the memory hierarchy. Fourth, the memory block that the main memory replied to the cache.
    
   The first chronograph illustrates the tile's execution with evenly distributed misses, resembling the baseline scenario. Consecutive warps demonstrate high locality, similar to the baseline. W\textsubscript{0} encounters a miss for block A in Cycle 1, followed by W\textsubscript{1}, W\textsubscript{2}, W\textsubscript{3}, and W\textsubscript{4} with secondary misses for block A over the next four cycles. Only one miss is served during these four cycles. Subsequently, W\textsubscript{5} encounters a miss for block B in Cycle 6. In Cycle 8, the cache receives a reply for block A from the memory hierarchy. As we progress, W\textsubscript{14} (the last warp) requests block C in Cycle 15 but receives its reply in Cycle 18, concluding its execution in cycle 18. Upon closer inspection, it is evident that the latencies of the misses do not efficiently overlap, thereby impacting the total execution cycles.

     Contrastingly, the second chronograph demonstrates the tile's execution when misses are clustered at the beginning by scheduling W\textsubscript{0}, W\textsubscript{5}, and W\textsubscript{10} at the beginning. Despite a 6X increase in hits, execution cycles increase to 21 due to the cache stalling for 6 cycles, after the MSHR becomes full at cycle 3.
     
     To strike a balance between memory parallelism and blocking, the last chronograph shows the tile's execution with clustered misses at the beginning, strategically spaced to prevent memory hierarchy blockages. W\textsubscript{5} follows W\textsubscript{0} to cluster misses, but W\textsubscript{10} is delayed until there is more space in the MSHR. This achieves an 11\% speedup compared to the baseline.
     
     Our objective with WaSP is to achieve a scenario akin to the one depicted in the last chronograph, where misses are clustered at the beginning of the execution but strategically spaced to prevent memory unit blockages. This tradeoff between memory parallelism and memory blocking is essential for optimizing the GPU core's performance in our proposal. While the additional 2 cycles in the initial example might seem inconsequential over an extended period, it's crucial to consider that TBR GPUs often encounter prolonged periods of low core occupancy. During these times, the available warps may not be sufficient to effectively mask long latencies.

     The rest of the section is divided into two subsections. One explaining how we selected the priority warps and the other explaining how we strategically spaced them to prevent memory hierarchy blockages.

%% file: 3_Technique/3_PW_selection.tex
\subsection{Priority Warps Selection}
    In the context of the aforementioned example, the question arises: How do we designate W\textsubscript{0}, W\textsubscript{5}, and W\textsubscript{10} as the priority subset at the time of launching the warp to the GPU core? Note that, at this stage, we lack information about the memory access addresses from these warps. Instead, we leverage the intrinsic properties of texture mapping and locality (as detailed in Section \ref{subsec:texlocality}) to make accurate predictions based on the screen coordinates of a quad (group of four pixels). This information is already employed by the baseline scheduler to allocate the warp to a specific GPU core \cite{dtexl}, where neighboring quads go to the same GPU core. We extend its use to choose the best set of priority warps.
    
    The goal is to identify a specific subset of warps that accesses the majority of texture blocks within the tile. It's crucial for this subset to be as compact as possible to avoid delayed prefetching for the regular warp subset and to maintain its locality-rich scanline order.

    To achieve this, we consider various subsets and evaluate two key ratios. Subset Size Ratio: the ratio of the number of warps in a subset to that of the whole tile. Texture Footprint Ratio: the ratio of the memory footprint of a subset to that of the entire tile. The goal is to compare these ratios and select the smallest subset that effectively represents the majority footprint of a tile. The following analysis aims to provide an approximate value for the ideal size of the subset we should target.

    \begin{figure}[t]
              \centering
              \includegraphics[width=\linewidth]{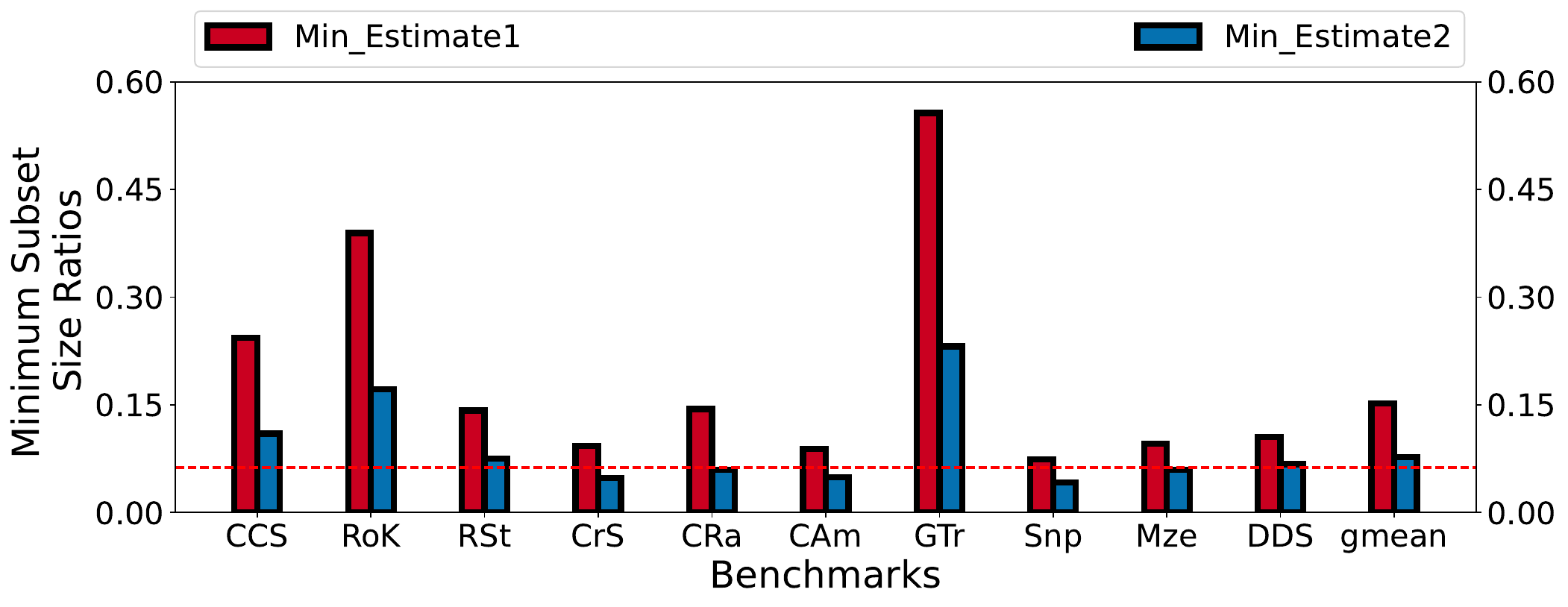}
              \caption{Estimation of a reasonable priority subset size.}
              \label{fig:Reuse_factor}
           \end{figure}

    \begin{figure*}[t!]
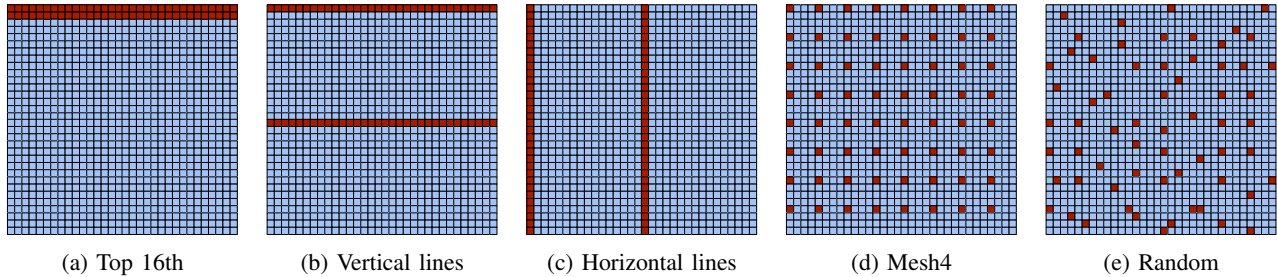

             \centering
             \begin{subfigure}{0.17\textwidth}
               \includegraphics[width=\linewidth]{3_Technique/Priority_subsets/Priority_top16th64.pdf}
               \caption{Top 16th}
               \label{fig:Pr-first4}
           \end{subfigure}
           ~
           \begin{subfigure}{0.17\textwidth}
               \includegraphics[width=\linewidth]{3_Technique/Priority_subsets/Priority_horz64.pdf}
               \caption{Vertical lines}
               \label{fig:Pr-horz}
           \end{subfigure}
           ~
           \begin{subfigure}{0.17\textwidth}
               \includegraphics[width=\linewidth]{3_Technique/Priority_subsets/Priority_vert64.pdf}
               \caption{Horizontal lines}
               \label{fig:Pr-vert}
           \end{subfigure}
           ~
           \begin{subfigure}{0.17\textwidth}
               \includegraphics[width=\linewidth]{3_Technique/Priority_subsets/Priority_Mesh4.pdf}
               \caption{Mesh4}
               \label{fig:Pr-Mesh2}
           \end{subfigure}
           ~
           \begin{subfigure}{0.17\textwidth}
               \includegraphics[width=\linewidth]{3_Technique/Priority_subsets/Mummy_Priority_Random64.pdf}
               \caption{Random}
               \label{fig:Pr-Random}
           \end{subfigure}

           \caption{Different Priority subsets.}
           \label{fig:Priority_Subsets}
        \end{figure*}

    \subsubsection{Subset Size Ratio}

           Before selecting a subset of warps as the priority subset, a crucial question arises: what should be the size of this subset? Priority warps, scheduled ahead of regular warps to emulate prefetching, must strike a balance in its subset size. Choosing too many priority warps can disrupt the inherent locality within the regular warps' order and diminish the benefits if their number is excessive. This effect is empirically illustrated in Section \ref{subsec:Meshesresults}. Thus, the size of the priority subset should be as small as possible.
           
           Determining the minimum size possible is not trivial and involves considering the characteristics of graphics applications. Here, we outline a rational approach that uses empirical data from our benchmark suite consisting of real-world graphics applications.
           
           Let's start with an assumption: each warp accesses only one memory block. Since blocks are reused and multiple warps can access the same memory block, the size of the smallest priority subset representing the entire tile's footprint equals the number of unique memory blocks accessed in that tile. Consequently, the subset size ratio can be calculated as:

           \[\text{Size ratio} = \frac{\# \text{Unique\_memory\_blocks\_in\_tile}}{\# \text{Warps\_in\_tile}}
           \]

           The red bar in Figure \ref{fig:Reuse_factor} represents this value for applications in our benchmark suite. The average value is 0.12. We call this the \textit{Min\_Estimate1}.
           
           However, warps can access multiple memory blocks based on the texels demanded and their corresponding mapped memory blocks. The priority subset can be made smaller in such cases. The best case is if we could find a warp that accesses all the memory blocks accessed in the tile. But such a warp is highly unlikely in real-world applications. Instead we empirically find the average unique memory blocks accessed per warp and name it `\textit{CF}'. We then assume each warp accesses these many blocks. Thus, as an example, if CF is 2, the minimum size of the priority subset becomes half of the red bar. This is a better estimate than \textit{Min\_Estimate1} and we call it \textit{Min\_Estimate2} and is presented below. Note that the empirical average value of CF across our benchmark suite is 2.5, which we later use it in our hueristic for WaSP, as explained in Section \ref{subsec:PW_scheduling}.
           
           Consequently, the size ratio becomes:
           \[
           \text{Size ratio} = \frac{\#\text{Unique\_memory\_blocks\_in\_tile}}{\#\text{Warps\_in\_tile} \times \text{CF}}
           \]                    

           The blue bar in Figure \ref{fig:Reuse_factor} represents \textit{Min\_Estimate2} for applications in our benchmark suite. This value is an approximation to the smallest possible subset size ratio for our applications. Hence, we choose the value 0.0625 (1/16), which is closest to the blue one. Accordingly, we select 1/16th of the total warps as our priority subset. Details on how these warps are chosen to effectively represent the footprint of a tile are elaborated in the next subsection.

    \begin{figure}[t]
              \centering
              \includegraphics[width=\linewidth]{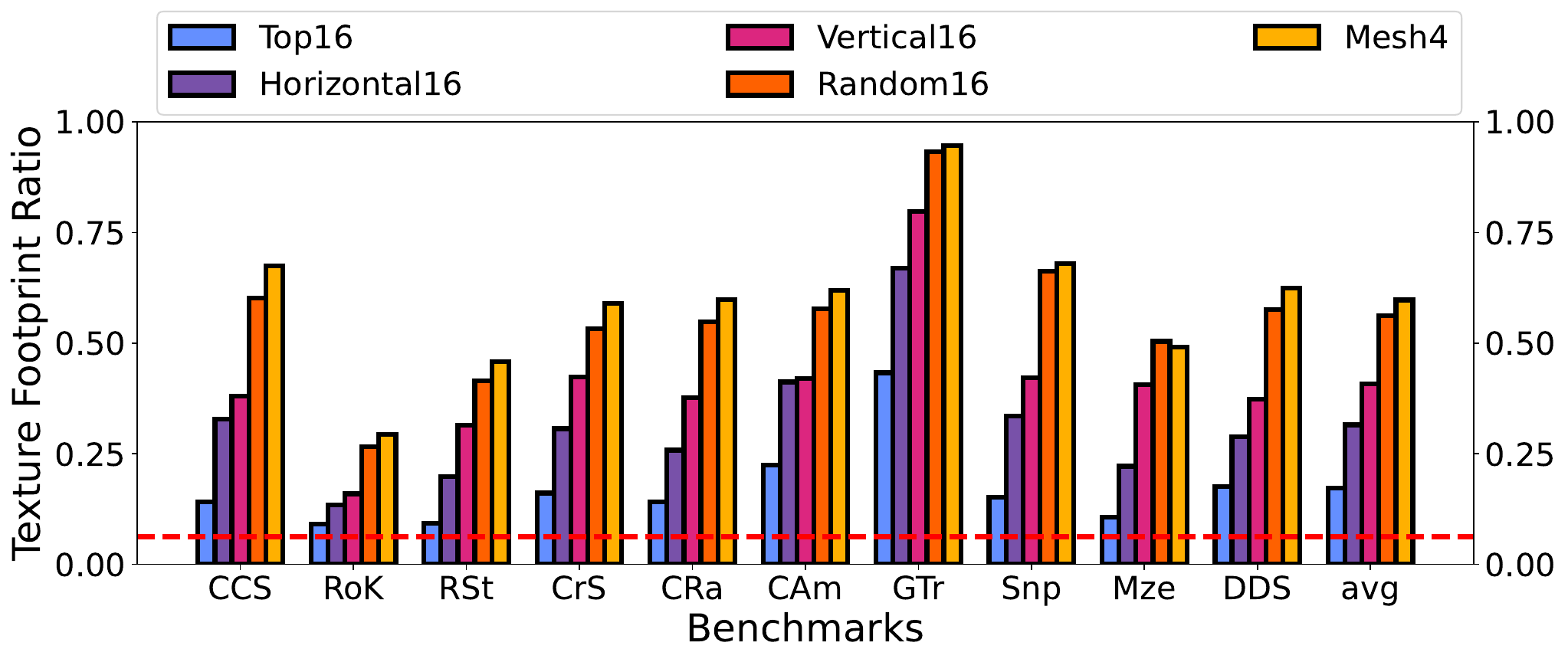}
              \caption{Fraction of texture footprint held by different subsets.}
              \label{fig:Frac_texfootprints}
    \end{figure}

       \subsubsection{Quad Selection}
           From Section \ref{subsec:texlocality}, we know that pixels located closely in proximity sometimes access the same texels, depending on their mapping to the texture plane and the type of texture filtering employed. In addition to temporal locality, textures are organized in a way that texels with spatial proximity are more likely to be stored in the same memory block. Therefore, when our objective is to select quads that access distinct texture memory blocks within a tile, aiming to pick the smallest subset that represents the largest possible texture footprint for the tile, it is intuitive to choose quads that are uniformly distributed spatially across the tile.
           
           Figure \ref{fig:Priority_Subsets} illustrates various methods of selecting the subset of priority quads, depicted in red, within a 32$\times$32 quad tile. These red quads indicate that the warps that map to these screen locations are priority warps. Figure \ref{fig:Frac_texfootprints} shows the ratio of the footprint of these subsets to the footprint of the tile. The red horizontal line indicates the value 0.0625 (1/16), representing the subset size ratio for all the subfigures.
           
           In Figure \ref{fig:Pr-first4}, the top one-sixteenth of the tile is chosen as priority warps, clustered together to access texture blocks overlapping only the top of the tile. The corresponding average footprint ratio, shown in Figure \ref{fig:Frac_texfootprints}, is 0.17, slightly higher than 0.0625, indicating a texture footprint ratio that exceeds the subset size ratio. This value is still low and thus this subset is not a good candidate to fully represent the entire tile footprint.
           
           In Figure \ref{fig:Pr-horz}, every 16th row of quads is chosen as a priority warp. This subset is clustered in the horizontal direction and distributed in the vertical direction. Thus it now accesses a higher number of texture blocks as the quads are farther away, the texture memory footprint will be higher. Figure \ref{fig:Pr-vert} shows the same but in vertical. We see that both of them show a better texture footprint ratio than the first (0.3 and 0.4 respectively). 
           
           Figure \ref{fig:Pr-Mesh2} partitions the tile into 4$\times$4 quad subtiles, selecting the top left corner as the priority warp. Consequently, the quads are distributed both horizontally and vertically. The resulting footprint ratio of 0.59 in Figure \ref{fig:Frac_texfootprints} signifies a close representation of the entire tile footprint. We call this subset selection \textit{Mesh4}. Mesh4 emerges as an excellent choice for priority warp selection in our proposal, aligning with our initial expectation that a more distributed subset of warps would encompass a variety of texture blocks in the priority subset. Note that by choosing just around 6\% of warps we have been able to represent almost 60\% of the full tile footprint. This can be attributed to the inherent locality in graphics applications.
           
           An alternative approach entails randomly choosing 1/16 of pixel locations within the tile as the priority warp subset. An illustration of this subset is presented in Figure \ref{fig:Pr-Random}, and the associated ratio in Figure \ref{fig:Frac_texfootprints} is 0.56. However, we favor Mesh4 for WaSP, given its cost-effectiveness (see Section \ref{subsec:hardwareoverhead}) and its capability to sufficiently represent the entire memory footprint.
           
           Therefore, Mesh4 is our proposed choice for WaSP.

%% file: 3_Technique/4_PW_timing.tex
    \subsection{Priority Warp Scheduling}
    \label{subsec:PW_scheduling}
        As previously outlined in section \ref{subsec:example}, the subset of priority warps is intentionally designed to have a higher density of unique memory block accesses. While this deliberate density aims to cluster misses to fully exploit memory-level parallelism, it also makes these priority warps more susceptible to memory unit blockages during the priority warp phase of tile execution. Scheduling all priority warps of a tile before the regular warps might prove counterproductive due to this blocking phenomenon.

        To address this issue, our solution involves initially scheduling just enough priority warps to the core, ensuring that the memory unit remains unblocked. Subsequently, regular warps are scheduled until the core has progressed sufficiently within the tile. This approach enables the core to handle additional cache misses without blocking the memory unit. However, this method presents two main challenges:

        \begin{itemize}
           \item The warp scheduler needs to anticipate future memory unit blockages at the time of launching a new warp into the GPU core, a task that is inherently complex.
           \item A priority warp must be scheduled sufficiently ahead of its adjacent regular warps to fulfill its intended prefetching purpose effectively.
        \end{itemize}

        \subsubsection{Blocking Prediction}
        \label{subsubsec:Block_prediction}
        To enable the warp scheduler to anticipate memory unit blocking, it broadly requires two pieces of information:

        \begin{itemize}
           \item The approximate number of long-latency misses that has been or will be generated by the warps currently in the GPU core.
           \item The status of the these misses, implying the count of misses already serviced, those currently being serviced, and those yet to be serviced.
        \end{itemize}

        To achieve this, we estimate the number of MSHR (Miss Status Handling Register) slots available in the L0 (for pipeline width 4) or L1 (for pipeline width 2) of a GPU core. In case the available slots are insufficient, this estimation prevents launching a new priority warp that could lead to misses, eventually filling all the MSHRs and consequently blocking the memory unit. A straightforward approach is to launch a new priority warp only when there are enough free slots in the MSHR. However, predicting how many currently executing warps will fill the MSHRs by the time the newly launched priority warp reached the LDST unit and encounters its own misses, is challenging. To calculate an approximate value for the number of misses that might be generated by the time a newly launched priority warp reaches the LDST (Load-Store) unit, we maintain a count of non-blocked priority warps in the GPU core.

        In GPUs, warps are marked as blocked when they encounter a long-latency miss, giving them lower priority for fetching instructions in the front-end. Each of the non-blocked priority warps has a high probability of generating CF number of misses. As mentioned before, the average value of CF obtained for our benchmark suite is 2.5. Thus we subtract these potential future misses from the available MSHR slots to efficiently approximate the \textit{Real\_freeMSHRs}.

        \begin{align*}
\text{Real\_freeMSHRs} = & \# \text{freeMSHRs} \\
& - (\# \text{nonblocked\_priority\_warps} \times 2.5)
\end{align*}
        
        If the calculated value exceeds a threshold, we launch a new priority warp. If it falls below the threshold, we instead launch a regular warp. Regular warps are less likely to cause misses unless launched before their neighboring priority warps. %

            In summary, we evaluate a boolean value (\textit{Priority\_over\_regular}) before initiating the launch of any warp into the GPU core. If the boolean evaluates to true, a priority warp (if available) is launched; otherwise, a regular warp is launched. Below, we present the final formula for this boolean condition.

            \begin{align*}
               \text{Priority\_over\_regular} &= (\text{Real\_freeMSHRs} > \text{Threshold} )
            \end{align*}

%% file: 3_Technique/5_HW_overhead.tex
   \subsection{ Hardware Overhead}
   \label{subsec:hardwareoverhead}
   The hardware overhead can be distinctly categorized for Priority Warp Selection and Priority Warp Timing.
   \subsubsection{Priority Warp Selection}
        After the Rasterizer, the warp scheduler allocates warps to the input FIFO queues of each GPU Core. WaSP divides this input queue into two distinct queues: the priority queue and the regular queue. Despite this division, the total size of the queues remains unchanged, avoiding any additional area overhead. WaSP then verifies whether a quad corresponds to the location of a priority warp before placing it into either queue. This verification is performed by checking if the \textit{x} and \textit{y} coordinates of the screen position of a quad are multiples of four, as priority warps are positioned at the top right corner of each subtile. This check involves examining the last two bits of the quad's coordinates. It's important to note that the baseline scheduler already evaluates these coordinates to assign warps to a specific GPU core. Therefore, the overhead incurred here is limited to a two-bit comparator per GPU core.

    \subsubsection{Priority Warp Scheduling}
        For warp scheduling, we require only two pieces of information from the GPU. The number of free MSHR slots of either the L0 or the L1 and the count of the total number of non-blocked warps in the core. This requires a 5-bit register to store the number of free MSHRs and 7-bit register to store the non-blocked priority warps count. We then require to calculate the boolean shown in the previous subsection using a multiplier and a comparator. Thus we see that the hardware overhead for WaSP is negligible compared to the total system cost.

%% file: 4_Methodology/Methodology.tex
\section{Evaluation Framework}
\label{sec:methodology}
    \begin{table*}[t]
           \begin{center}
              \caption{Evaluated benchmarks from the Google Play Store.}
              \label{tab:benchmarks}
              \begin{tabular}
              {llrlcrr}
                 \hline
                 \textbf{Benchmark} & \textbf{Alias} &
                 \thead[r]{\bf Installs \\ \bf (Millions)} &
                 \textbf{Genre} & \textbf{Type}& 
                 \thead[r]{\bf Memory Block \\ \bf reuse in tile} & 
                 \thead[r]{\bf Memory-Instr ratio } \\
                 \hline
                 \rowcolor{lightgray}
                 Candy Crush Saga & CCS & 1000 & Puzzle & 2D & 8.1 & 0.25\\
                 Rise of Kingdoms: Lost Crusade& RoK & 10 & Strategy & 2D & 4.8 & 0.08\\                 
                 \rowcolor{lightgray}
                 Real Steel World Robot Boxing & RSt & 50 & Strategy & 3D & 12.3 & 0.15\\
                 Crazy Snowboard & CrS & 5 & Sports & 3D & 19.8 & 0.18\\
                 \rowcolor{lightgray}
                 City Racing 3D& CRa & 50 & Racing & 3D & 15.8 & 0.15\\
                 Captain America & CAm & 5 & Action & 3D & 19.3 & 0.26\\
                 \rowcolor{lightgray}
                 Gravitytetris & GTr & 5 & Puzzle & 3D & 3.3 & 0.09\\
                 
                 Sniper 3D & Snp & 500 & Shooter & 3D & 22.9 & 0.15\\
                 \rowcolor{lightgray}
                 3D Maze 2: Diamonds \& Ghosts & Mze & 10   & Arcade & 3D & 15.8 & 0.12\\
                 Derby Destruction Simulator & DDS & 10 & Racing & 3D & 13.9 & 0.07\\

                 \hline
               \end{tabular}
            \end{center}
        \end{table*}

        \begin{table}[t]
           \begin{center}
              \caption{GPU simulation parameters.}
              \label{tab:Simulation Parameters}
                 \begin{tabular}{ll}
                    \hline
                    \multicolumn{2}{c}{\textbf{Global Parameters}}\\
                    \hline
                    Tech Specs & 600MHz, 1V, 32nm\\
                    Screen Resolution & 1960$\times$768\\
                    Tile Size & 64x64\\
                    Tile Traversal Order & Z-order\\
                    \hline
                    \multicolumn{2}{c}{\textbf{Main Memory}}\\
                    \hline
                    Latency & 50-100 cycles\\
                    Size & 1GiB\\
                    \hline
                    \multicolumn{2}{c}{\textbf{Caches}}\\
                    \hline
                    Vertex Cache & 64-bytes/line, 64KiB, 4-way, 1 cycle\\
                    Texture Caches (4x) & 64-bytes/line, 64KiB, 4-way, 1 cycle\\
                    Tile Cache & 64-bytes/line, 64KiB, 4-way, 1 cycle\\
                    L2 Cache & 64-bytes/line, 1MiB, 8-way, 18 cycles\\
                    \hline
                  \end{tabular}
                \end{center}
        \end{table}
    \subsection{GPU Simulation Framework}
    
    We use the TEAPOT~\cite{Teapot} simulation infrastructure to evaluate our proposal. TEAPOT is a cycle-accurate GPU simulation framework that allows to run unmodified Android applications and evaluates the performance and energy consumption of the modeled GPU. In order to do that, TEAPOT includes timing and power models based on well-known tools: McPAT~\cite{McPaT} for power estimation, and DRAMSim2~\cite{DRAMSim2} for modeling DRAM and the memory controllers. Table~\ref{tab:Simulation Parameters} shows the parameters employed in our simulations, which resemble those of a contemporary mobile GPU.

    \subsection{Benchmarks}
    
        We use popular commercial animated graphics applications (games) as benchmarks. We have selected them based on their popularity, in the number of downloads in the Google Play Store, and their variety to cover different types of games.
        
        Table~\ref{tab:benchmarks} shows the twelve Android games used to evaluate our technique. We have 2D games like $CCS$ and 3D games like $CRa$. Some characteristics of these games can give us an insight into how well it will work with WaSP. Table~\ref{tab:benchmarks} shows that $RoK$ and $GTr$ low reuses of texture blocks and a low memory instruction ratio in their kernels. A high reuse implies that prefetched memory blocks will benefit more number of warps. And a high memory-instruction ratio implies that stalls in the memory unit are memory-bound and have a higher chance of affecting performance. Thus $RoK$ and $GTr$ should not work very well with WaSP. While this is true for $RoK$, it is not true for $GTr$. This is because Mesh4 gives a low texture footprint ratio for $RoK$ compared to $GTr$, as shown in Figure \ref{fig:Frac_texfootprints}.

%% file: 5_Evaluation/Evaluation.tex
\section{Experimental Results}
\label{sec:results}
    In this Section, we present empirical evidence validating the efficacy of WaSP. This Section is divided into three parts: the first part is a sensitivity analysis on the pipeline width and the maximum warps allowed in the tile. The next part assesses WaSP for a pipeline width of 4 and maximum warps of 64. We describe our parameter tuning process outlined in Section \ref{subsec:PW_scheduling} and present outcomes based on the most optimal parameter set.
    The last part assesses the efficacy of our priority warp selection for WaSP with varying subset size ratios for the priority warp set. The parameters of the GPU assumed are outlined in Table \ref{tab:Simulation Parameters}.
    \begin{figure*}[t!]
             \centering
             \begin{subfigure}{0.4\textwidth}
               \includegraphics[width=\linewidth, trim=6.1cm 4cm 5cm 4cm, clip]{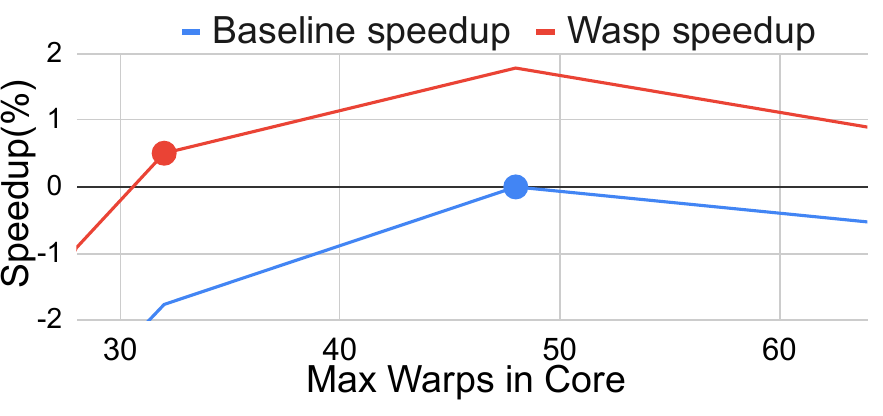}
               \caption{ Speedup w.r.t. Optimal baseline 48warps (Pipeline Instruction Width 2). }
               \label{fig:PW2Sense}
           \end{subfigure}
           \hspace{0.1\textwidth} %
           \begin{subfigure}{0.4\textwidth}
               \includegraphics[width=\linewidth, trim=6.1cm 4cm 5cm 4cm, clip]{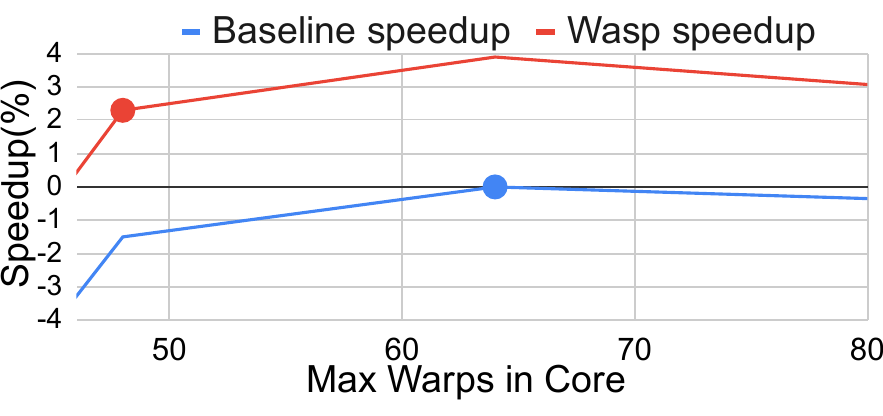}
               \caption{Speedup w.r.t. Optimal baseline 48warps (Pipeline Instruction Width 4).}
               \label{fig:PW4Sense}
    \end{subfigure}

           \caption{Sensitivity Analysis.}
           \label{fig:sensitivity}
        \end{figure*}

    \subsection{Sensitivity Analysis}
        Let us revisit an earlier point: GPUs employ multithreading to hide latency. The efficiency of hiding latency relies on the number of warps needed to fill the pipeline's width and depth. Contemporary GPU cores typically feature 2-instruction-wide pipelines. However, future advancements might introduce wider pipelines to enhance instruction parallelism within a core. Consequently, the need for additional warps in the core to manage latency and sustain performance increases. However, expanding the number of warps incurs higher costs in terms of area and energy consumed by core structures, notably the register file, which significantly contributes to the overall GPU area. This is where WaSP can offer assistance.
        
        In this subsection, we conduct a sensitivity analysis to assess WaSP's performance improvement while varying two crucial parameters: the pipeline width (with values of 2 and 4) and the maximum permissible number of warps within a GPU core (with values of 28, 32, 48, 64, 128). 
        
        Figure \ref{fig:PW2Sense} is for a pipeline width of 2. Here, the x-axis varies the maximum warps allowed in a GPU, whereas the y-axis shows the speedup of the baseline and WaSP with w.r.t. the baseline of the optimal number of warps. In this case, the optimal number is 48. The blue line shows the speedup of the baseline w.r.t. the optimal number. The slow down after 48 warps shows that there is resource contention if we fill the core more. The red line, which is for WaSP, is around 2\% better than the blue line, which is for the baseline. Note that the highlighted points show that WaSP with just 32 warps can outperform the optimal baseline with 48 warps.

        Figure \ref{fig:PW4Sense} is for a pipeline width of 4. It is clear to see that this is a scaled and shifted version of the previous figure. Now, the optimal number of warps shifts to 64 due to the increased pipeline width. The benefits of WaSP also double, achieving a speedup of 4\%. Also, the highlighted points show that WaSP with just 48 warps performs 2\% better than the optimal baseline with 64 warps. This implies a reduction of one-fourth of the register file area, improving performance at the negligible cost of WaSP.

    \subsection{WaSP w.r.t. the best baseline}
        Based on the preceding analysis, our findings indicate that the most optimal baseline configuration entails a pipeline width of 4 with a maximum of 64 warps in the core. In this Section, we delve into an intricate examination of WaSP concerning this optimal baseline.
        
        WaSP's goal revolves around early scheduling of the priority warps, aiming to reduce latency for regular warps without compromising the inherent locality maintained by the baseline scheduler. We present a range of metrics below to validate this hypothesis regarding WaSP's effectiveness. We vary some parameters in the scheduling heuristic to explain how we fine tuned the parameters to achieve the best results. We also explain the variations across applications corresponding to their characteristics detailed in Table \ref{tab:benchmarks}. Note that the applications are arranged in the decreasing order of their ideal speedup as shown in Figure \ref{fig:motivation}.

        \begin{figure}[t]
              \centering
              \includegraphics[width=\linewidth]{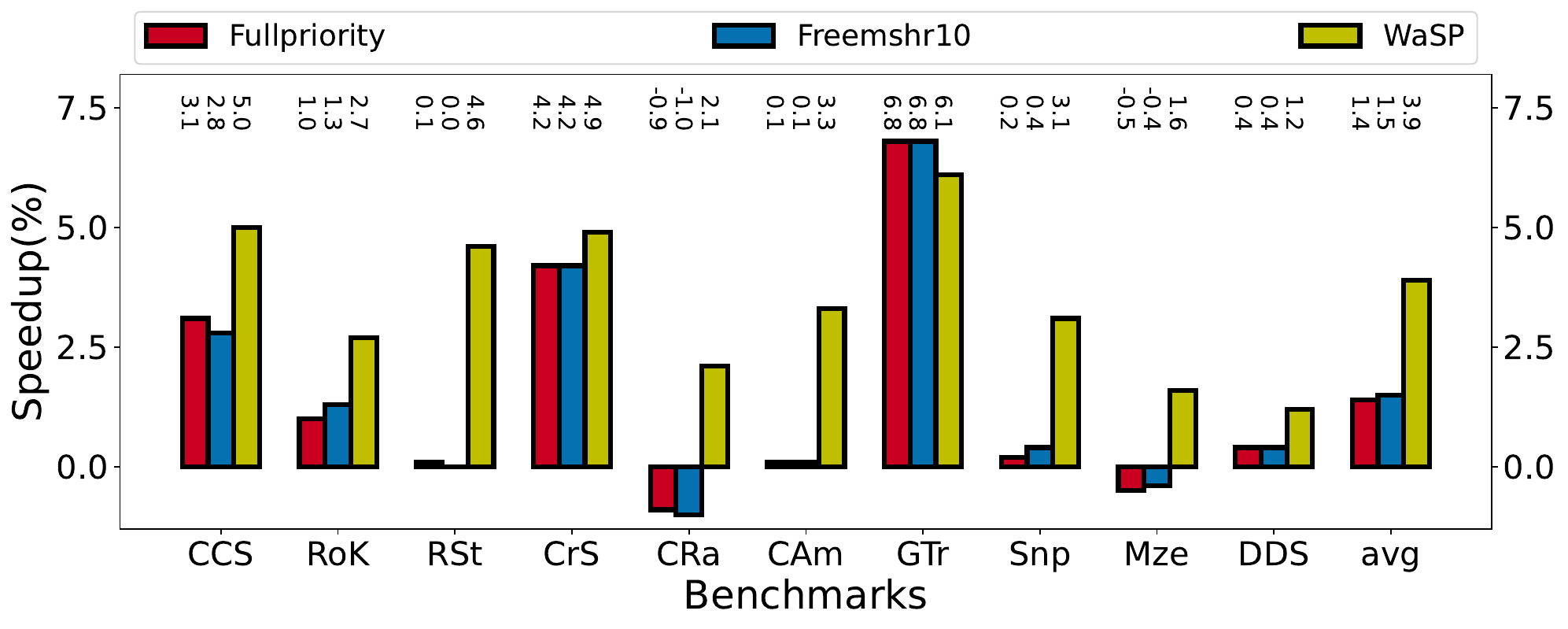}
              \caption{Speedup with different variations of WaSP.}
              \label{fig:PWsched_IPC}
        \end{figure}

        \begin{figure}[t]
              \centering
              \includegraphics[width=\linewidth]{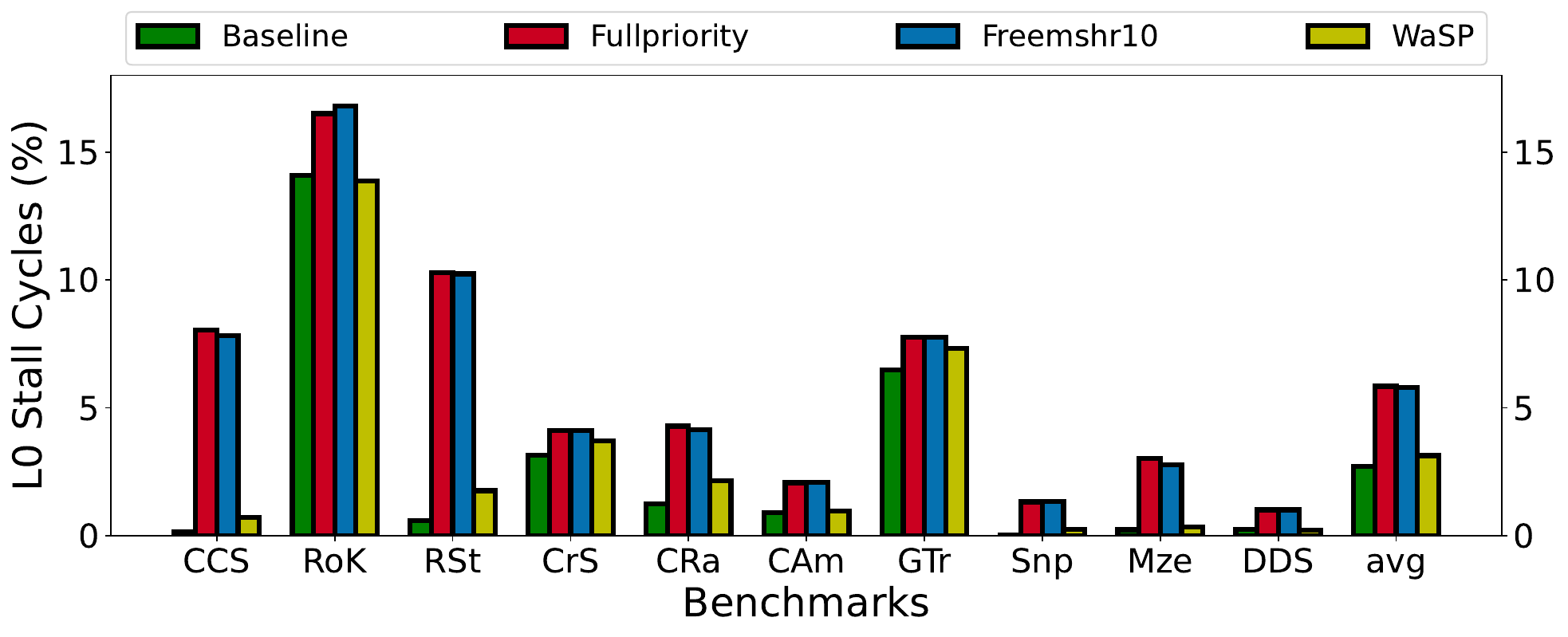}
              \caption{Percentage of execution time L0 is stalled.}
              \label{fig:L0Stall}
        \end{figure}
        
        \subsubsection{Performance}
            Figure \ref{fig:PWsched_IPC} displays the speedup achieved by various heuristics, including WaSP. The \textit{Fullpriority} heuristic involves scheduling all priority warps ahead of regular warps. Despite an average speedup of 1.3\%, notable variation exists across applications, with $CRa$ experiencing a slowdown. This trend aligns with Figure \ref{fig:L0Stall}, indicating the percentage of time the L0 of a core is stalled w.r.t. execution time. Specifically, $CrS$ and $GTr$ show minimal changes in stall time compared to the baseline, while other applications demonstrate significant alterations. We see that $CrS$ and $GTr$ work reasonably well with \textit{Fullpriority}. This increased stall time for the rest of the applications is in line with our explanation in Section \ref{subsec:example}. Thus  we explore scheduling priority warps while minimizing the stalling of the cache.

            The \textit{Freemshr10} heuristic schedules priority warps into the core only when at least 10 MSHRs are available. As explained in Section \ref{subsec:PW_scheduling}, this does not take into account the number of MSHRs that will be occupied by the time the launched warp reaches the LDST unit. Thus we see a similar behavior to \textit{Fullpriority}.

            Subsequently, employing WaSP yields a 3.9\% speedup by considering the MSHRs occupied by other priority warps in the GPU core by the time the currently launched warp reaches the LDST unit. This speedup aligns with the decreasing order of speedup depicted in Figure \ref{fig:motivation}, barring two outliers: $RoK$ and $GTr$. $RoK$ displays the lowest texture footprint ratio for the priority subset, impacting its performance compared to others. Conversely, $GTr$, with a texture footprint ratio close to one, exhibits exceptional performance with WaSP. This observation is illustrated in Figure \ref{fig:Frac_texfootprints}.

        \begin{figure}[t]
              \centering
              \includegraphics[width=\linewidth]{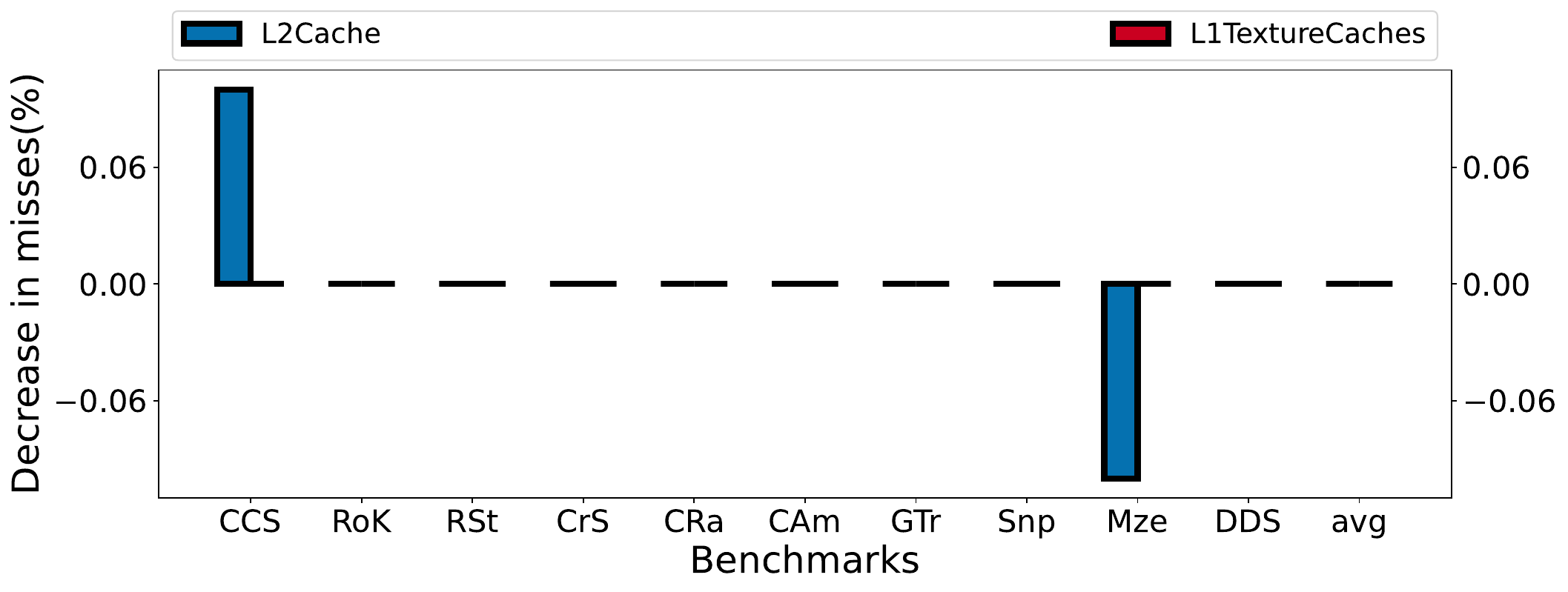}
              \caption{ Decrease in L1 and L2 misses wrt baseline.}
              \label{fig:PWsched_L1L2misses}
        \end{figure}

        \subsubsection{Misses of L1 and L2}
            To validate our hypothesis that WaSP preserves locality, we initially examine the alterations in L1 and L2 cache misses. Figure \ref{fig:PWsched_L1L2misses} illustrates the reduction in both L1 and L2 misses attributed to WaSP. Notably, there is negligible variance in misses observed in both L1 and L2 caches. This substantiates our earlier assumption outlined in Section \ref{subsec:example} when conceptualizing WaSP. We theorized that once a memory block enters L1 and L2 caches, it remains there until the tile's execution concludes. Consequently, the working set of a tile comfortably fits within L1 and L2 caches, affirming that WaSP does not compromise the inherent locality maintained by the baseline scheduler.

        \begin{figure}[t]
              \centering
              \includegraphics[width=\linewidth]{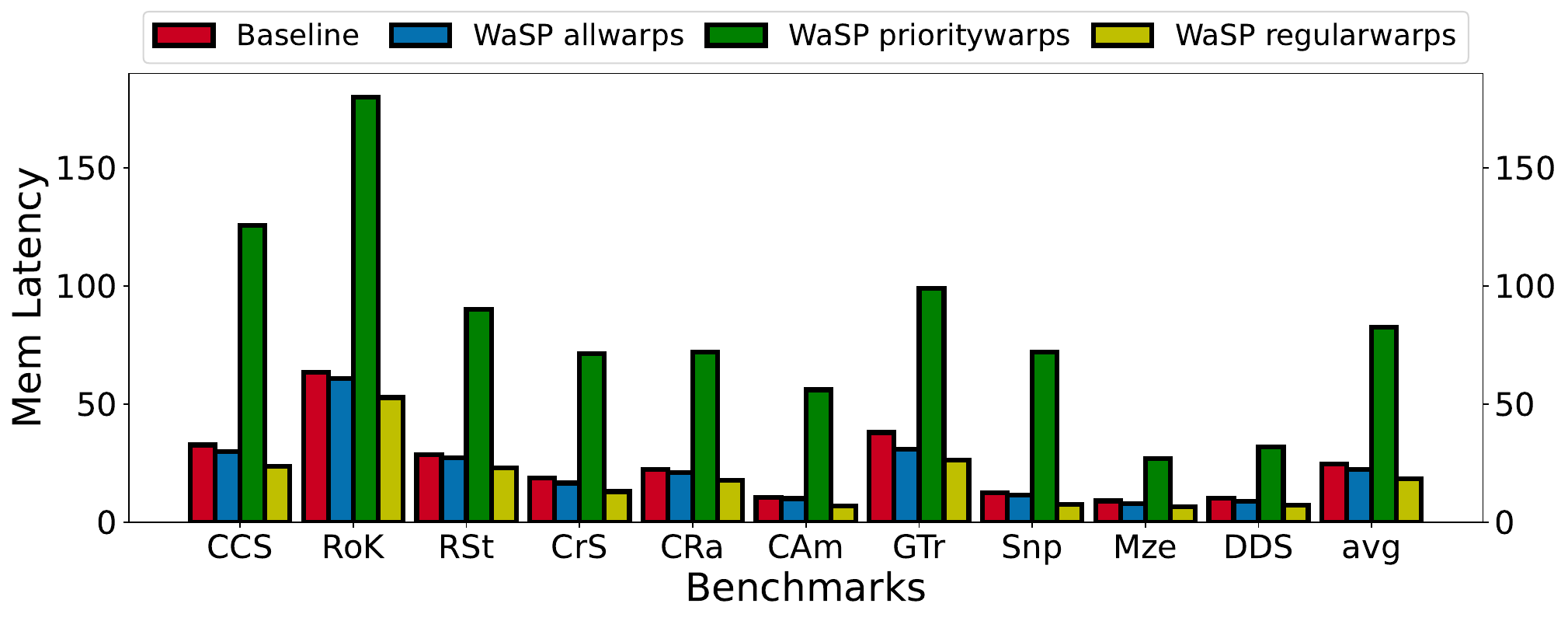}
              \caption{Average memory latency of warps.}
              \label{fig:PWsched_memlatency}
        \end{figure}
        \subsubsection{Memory Latency}
            Figure \ref{fig:PWsched_memlatency} showcases the average reduction in memory latency across all applications within the benchmark suite. A noteworthy average decrease of 9\% in memory latency is evident without any concurrent change in the count of L1 and L2 misses. This suggests that the reduction may stem from either a decreased latency of secondary misses or the transformation of secondary misses into hits. These outcomes are a direct result of WaSP's prefetch emulation, showcasing its ability to emulate prefetching and harness memory parallelism at beginning of a tile. Notably, $GTr$ demonstrates a reduction of around one-fifth in average memory latency. This aligns with Figure \ref{fig:Frac_texfootprints}, where $GTr$ exhibits the highest texture footprint ratio of 0.9, indicating that priority warps prefetch for 90\% of the texture footprint within the tile.

            For WaSP we also split the average memory latency for the priority subset and the regular subset. The average memory latency for priority warps notably exceeds that of regular warps, reaffirming our initial assumptions regarding WaSP's behavior and priorities.

        \begin{figure}[t]
              \centering
              \includegraphics[width=\linewidth]{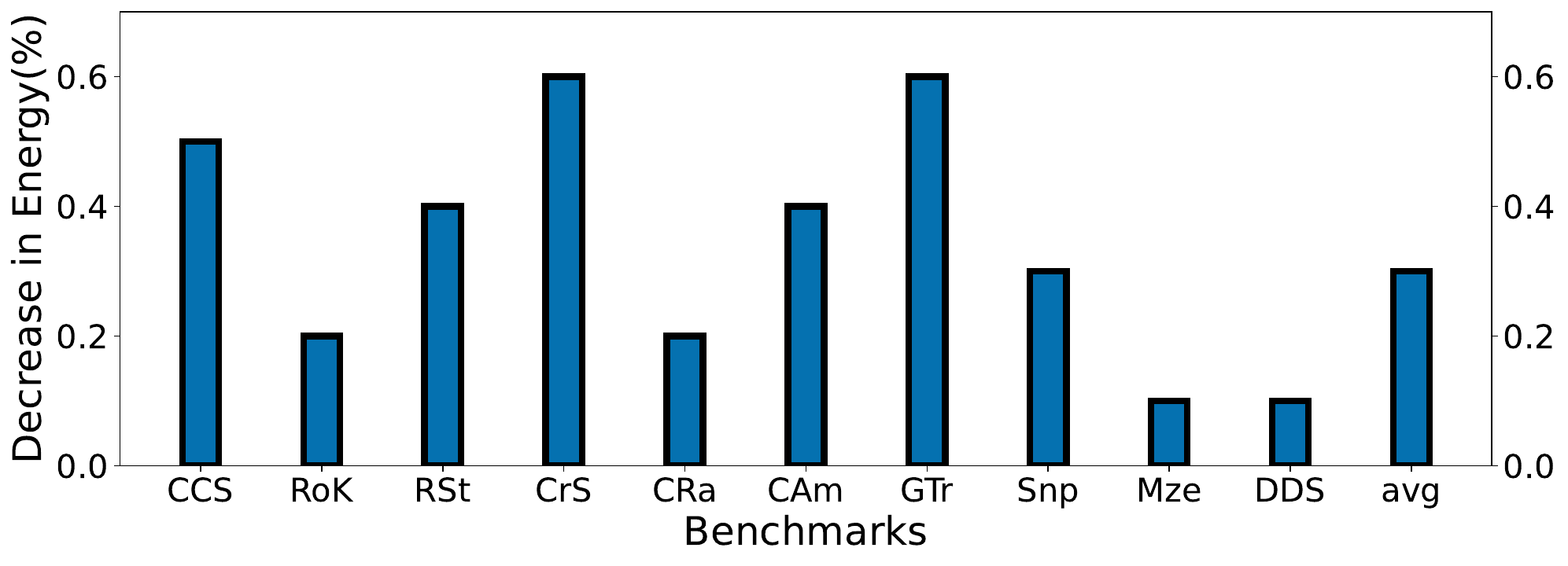}
              \caption{Decrease in total GPU Energy(\%).}
              \label{fig:Energy}
        \end{figure}
        
        \subsubsection{Energy}
            Figure \ref{fig:Energy} shows the decrease in full system energy of WaSP w.r.t. the baseline. We see a decrease of 0.5\%. This correlates with the decrease in execution time. Remember that although the numbers seem small this 3.9\% gain in performance and 0.5\% drop in energy comes with a negligible cost.

    \begin{figure}[t]
              \centering
              \includegraphics[width=\linewidth]{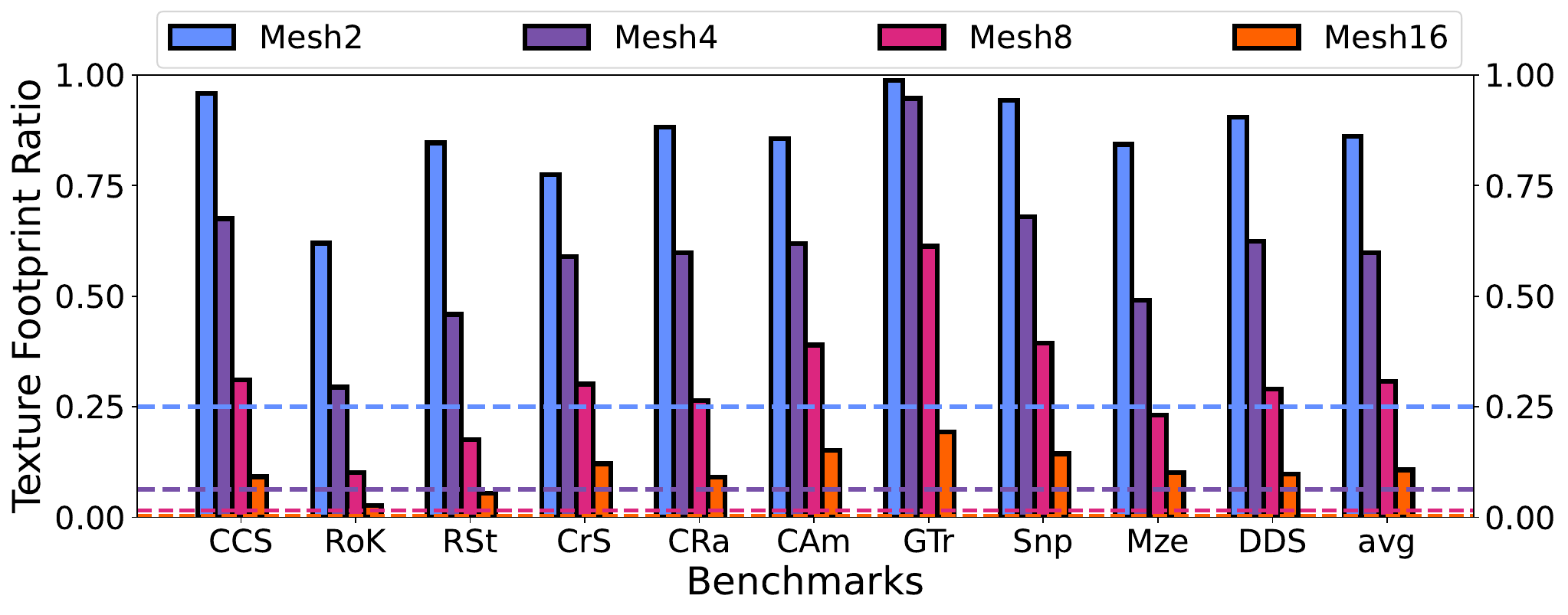}
              \caption{Fraction of texture footprint held by different subsets with varying mesh size. The horizontal lines indicate the corresponding subset size ratios.}
              \label{fig:Frac_texfootprintsmeshes}
    \end{figure}

    \begin{figure}[t]
              \centering
              \includegraphics[width=\linewidth]{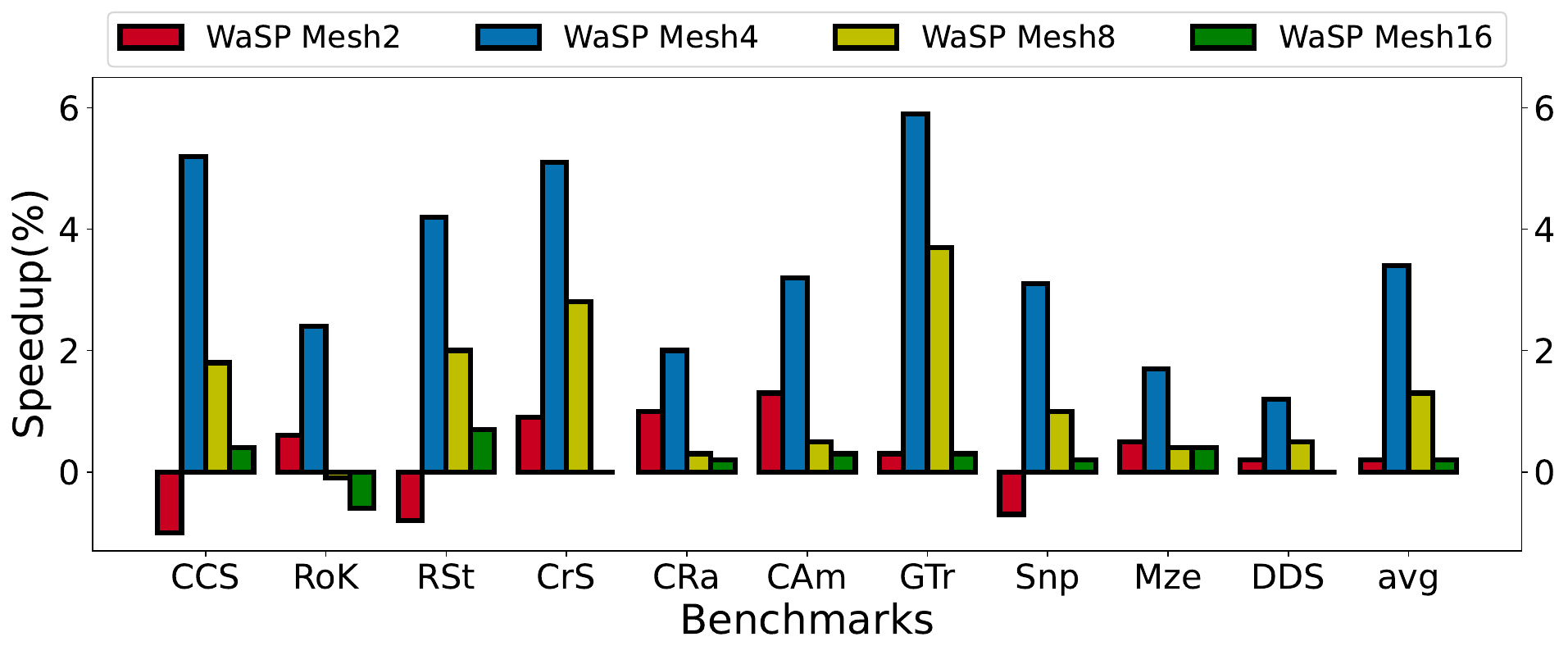}
              \caption{Speedup of different subset size ratios.}
              \label{fig:Frac_meshspeedup}
    \end{figure}

    \subsection{Priority Warp Selection}
        \label{subsec:Meshesresults}
        Figure \ref{fig:Frac_texfootprintsmeshes} portrays the Texture Footprint Ratio which is the ratio of the footprint of the priority warp subset to that of the complete set of warps within a tile. We explore priority subsets with varying Subtile Size Ratios. Subtile Size Ratio is the ratio of the total number of warps in a priority subset to that in the whole tile. When using a subtile size of 2$\times$2 quads using uniform distribution as in Figure \ref{fig:Pr-Mesh2} (which we call Mesh2), we observe an average texture footprint ratio of 0.9, which provides a very effective coverage of the tile's footprint. However, using 2$\times$2 quad subtiles results in a high subset size ratio of 0.25, as depicted by the corresponding horizontal line in the same color, because we are choosing one priority warp out of four quads. This notably impacts warp locality as shown in Figure \ref{fig:Frac_meshspeedup}, which shows the speedup of WaSP for different priority sets. For Mesh2, there is a slowdown for three applications and the absence of significant speedups for others. 
        
        In contrast, employing a subtile size of 4$\times$4 quads, as we previously have in the rest of Section \ref{sec:results}, yields an average ratio of 0.64 with a substantially smaller subset size ratio of 0.0625. As shown in Figure \ref{fig:Frac_meshspeedup}, this configuration emerges as the most optimal choice. Additionally, we found that smaller subsets, such as 8$\times$8 and 16$\times$16, are feasible but exhibit significantly lower texture footprint ratios and speedups. This reinforces our selection of Mesh4 as the optimal configuration for WaSP.

%% file: 6_Related_Work/Related_work.tex
\section{Related Work}
\label{sec:related_work}

Previous GPU warp scheduling studies prioritized distributing warps among GPU cores to enhance load balancing or locality, without adjusting the sequence in which each core obtains its assigned warps. Despite various investigations into scheduling and prefetching, none, like WaSP, have specifically recommended scheduling strategies that emulate prefetching techniques.

Kerbl et al.'s work~\cite{Load_balancing} introduced a static workload scheduler for parallel tile rendering architectures with multiple Raster Units. Their focus was on distributing tiles among these units, aiming for load balancing by ensuring an even thread distribution across Raster Units. Other studies like \cite{fuchs1989pixel} and \cite{dynamiclin} also explore dynamic workload scheduling in similar architectures but don't reorder the warps but rather distribute the workload amongst cores. 

Ukarande et al. \cite{Nvidia_tex_locality} emphasized the significance of locality-driven workload scheduling, achieving a 4\% speedup in high-end desktop graphics workloads by leveraging Texture Cache locality. Their method involved clustering Cooperative Thread Arrays (CTAs) based on proximity in screen coordinates. Similar approaches in other studies such as \cite{Texlocality_CTAcluster}, \cite{CTA_cluster_GPGPU}, and \cite{Threadblockscheduling_GPGPU} focused on CTA scheduling for L1 Data Cache locality in GPGPU workloads, yielding notable performance enhancements. Notably, these approaches are purely software-based and again focus on assignment and not sequence, whereas WaSP is a hardware warp scheduler which changes the sequence of warp assignment within a core. 

Several studies have delved into enhancing cache locality within GPUs for GPGPU workloads. Some have focused on cache locality spanning kernel launches for related kernels such as parent-child kernels \cite{Locality_aware_scheduling_GPGPU} or interdependent kernels \cite{Locality_aware_scheduling_GPGPU2}. Additionally, research like \cite{14}, \cite{Cache_Throttling1}, and \cite{Texlocality_CTAcluster} has investigated the effects of warp throttling on locality in GPGPU workloads. Others, such as \cite{Warp_scheduling1}, \cite{Warp_scheduling2}, \cite{Warp_scheduling3}, \cite{Warp_scheduling4}, \cite{Warp_scheduling5}, and \cite{OWL}, have explored warp-scheduling impact within a core on locality for GPGPU workloads. Furthermore, studies like \cite{21}, \cite{25}, \cite{26}, \cite{47}, \cite{56}, \cite{58}, \cite{Cache_bypassing1}, \cite{Cache_bypassing2}, \cite{Cache_bypassing3}, \cite{Cache_bypassing4}, and \cite{Cache_bypassing5} have delved into cache bypassing techniques to bolster GPU cache locality. Another approach presented in \cite{11} suggests implementing a Cooperative Caching Network (CCN) - a ring network linking L1 caches within a GPU - aiming to alleviate L2 bandwidth demand.

Hardware prefetching techniques in GPUs have been previously explored. One of the approaches \cite{igehy1998prefetching, Jose_Maria1} is to calculate the texture addresses ahead of scheduling any warp into the GPU core. This has a hardware overhead. For GPGPU workloads APOGEE\cite{sethia2013apogee} proposes a dynamically adaptive prefetching technique. Other GPGPU prefetchers are \cite{falahati2015powerprefetch, prefetch1, prefetch2, prefetch3, prefetch4, prefetch5, prefetch6}. None of the above techniques propose prefetching using warp scheduling.

Other recent works with graphics workloads have explored memory bandwidth reduction in TBR architectures using various methods. Early Visibility Resolution (EVR) \cite{EVR} is an HSR technique that speculatively predicts the visibility of objects in a scene before the Raster Pipeline to avoid computation and texture accesses of fragments that will eventually be discarded. Rendering Elimination \cite{Render_elimination} is a technique that detects tiles that produce the same color across adjacent frames to avoid redundant computation and texture accesses. Another work, TCOR \cite{TCOR}, explores memory bandwidth reduction by targeting another major source of main memory accesses in TBR architectures, which is the Parameter Buffer.

All these methods remain orthogonal to WaSP, a framework that focuses on reducing cache latency by overlapping more misses through a warp reordering strategy while ensuring the preservation of inherent locality.

%% file: 7_Conclusions/Conclusions.tex
\section{Conclusions}
\label{sec:conclusions}
In this work we introduce WaSP, a lightweight warp scheduler for GPUs for graphics applications that emulates prefetching a tile's working set early in the tile's execution, aiming to minimize memory latency for subsequent warps within the tile. This is done by harnessing the existing underutilized memory parallelism at the beginning of the tile without blocking the memory unit, ensuring that performance is not only maintained but also enhanced in the process.

WaSP achieves this by segregating a subset of warps that accurately represents the majority of texture accesses in the tile, thereby optimizing memory parallelism. Priority warps are scheduled into a GPU core until the memory unit is close to saturation due to outstanding misses. At this point, regular warps are scheduled, ensuring a more efficient use of the memory resources and reducing the average miss latency for memory accesses in all warps.

Multi-threading is generally used in GPUs to hide these long latencies. But increasing the number of warps has a high cost in hardware. WaSP is able to achieve the same speedups as a significant increase in the number of warps. For example, WaSP with 32 warps is able to achieve the same performance as the baseline with 48 warps (1.5 $\times$ register file size compared to 32 warps).

In summary, this paper presents several significant contributions to the field. Firstly, it introduces the concept of Subset Optimization, which involves proposing and assessing different warp subsets as priority warps to determine the best representation for a tile's majority texture footprint. Secondly, the paper discusses scheduling heuristics, presenting and evaluating two strategies for transitioning between priority and regular warps. These heuristics effectively prevent memory unit blocking due to priority warps, leading to an overall performance improvement. Lastly, the paper demonstrates notable performance improvements, showcasing a 3.9\% speedup, almost for free. These outcomes emphasize the efficiency gains achieved through the WaSP scheduler.